# Vestibular Drop Attacks and Meniere's Disease as Results of Otolithic Membrane Damage – A Numerical Model

Nicholas Senofsky[1,a,A], Justin Faber[1,b,A], and Dolores Bozovic[1,2,c,A]

1) Department of Physics and Astronomy, University of California, Los Angeles, California 90095, USA
2) California NanoSystems Institute, University of California, Los Angeles, California 90095, USA
a) nicksenofsky@ucla.edu
b) faber@physics.ucla.edu
c) bozovic@physics.ucla.edu
A) A-432 Physics and Astronomy Building
   430 Portola Pl
   Los Angeles, CA 90095-1547
   (310) 267-5881

**ABSTRACT:**

**BACKGROUND:** Meniere's disease (MD) is a condition of the inner ear with symptoms affecting both vestibular and hearing functions. Some patients with MD experience vestibular drop attacks (VDAs), which are violent falls caused by spurious vestibular signals from the utricle and/or saccule. Recent surgical work has shown that patients who experience VDAs also show disrupted utricular otolithic membranes.

**OBJECTIVE:** The objective of this study is to determine if otolithic membrane damage alone is sufficient to induce spurious vestibular signals, thus potentially eliciting VDAs and the vestibular dysfunction seen in patients with MD.

**METHODS:** We use a previously developed numerical model to describe the nonlinear dynamics of an array of active, elastically coupled hair cells. We then reduce the coupling strength of a selected region of the membrane to model the effects of tissue damage.

**RESULTS:** As we reduce the coupling strength, we observe large and abrupt spikes in hair bundle position. As bundle displacements from the equilibrium position have been shown to lead to depolarization of the hair-cell soma and hence trigger neural activity, this spontaneous activity could elicit false detection of a vestibular signal.

**CONCLUSIONS:** The results of this numerical model suggest that otolithic membrane damage alone may be sufficient to induce VDAs and the vestibular dysfunction seen in patients with MD. Future experimental work is needed to confirm these results *in vitro*.

**INTRODUCTION:**

Meniere's Disease (MD) is a condition of the inner ear affecting both vestibular and hearing functions. MD occurs in 17 people out of 100,000, and adequate treatment options are scarce. In some cases, clinicians resort to gentamicin profusions, vestibular neurectomies, or even labyrinthectomies. MD can cause episodic vertigo, hearing loss, tinnitus, and excessive pressure in the inner ear, with symptoms occurring either unilaterally or bilaterally [17]. In rare cases, patients affected by MD can suffer from vestibular drop attacks (VDAs), which are characterized by a sudden, forceful loss of balance. The cause of vestibular drop attacks is currently unknown. However, recent work has suggested that the presence of VDAs among patients with MD can be attributed to damage of the otolithic membrane tissue in the utricle and/or saccule [2]. Calzada found that all patients who experience VDAs show disrupted utricular otolithic membranes. Additionally, Calzada showed that the average thickness of the utricular otolithic membrane in patients with MD is 11.45 micrometers, while in normal tissue the mean thickness is 38 micrometers. Patients often report multiple vestibular drop attacks, with the resulting falls occurring in the same direction each time, suggesting a spurious vestibular signal. Meniere's Disease is highly correlated with endolymphatic hydrops. However, the two conditions are clinically distinct from each other, and the mechanisms causing MD are currently unknown [5].

The utricle and saccule rely on hair cells to detect linear accelerations. These specialized sensory cells are named after the rodlike stereovilli that protrude from their apical surface. The stereovilli are collectively named the hair bundle, which pivots in response to acceleration. This deflection of the hair bundle modulates the tension in the tip links that connect adjacent rows of stereovilli and controls the gating of mechanotransduction channels embedded in the tips of the stereovilli. Therefore, hair cells transduce the mechanical energy of acceleration into electrical energy in the form of ionic currents into the cell [7, 13, 19]. The response of the inner-ear hair cells to incoming signals has been shown to be highly nonlinear, in a compressive manner, allowing them to exhibit a broad dynamic range [11]. Further, hair cells produce active amplification, enhancing weak signals and thus endowing auditory and vestibular systems with high sensitivity [9].

Hair-cell bundles of certain species have been shown to exhibit innate motility in the absence of stimulus [1, 8]. In the amphibian sacculus, these spontaneous oscillations have amplitudes significantly larger than the motion induced by thermal fluctuations in the surrounding fluid. These innate oscillations have been shown to be active, powered by an energy-consuming process [10]. While they have been proposed to assist in the amplification of weak signals, or possibly underlie the generation of otoacoustic emissions by the inner ear, the presence or role of these spontaneous oscillations *in vivo* is currently unknown. Experiments performed *in vitro* on the amphibian sacculus, however, have shown that these active oscillations are suppressed by the presence of the overlying otolithic membrane [4]. In healthy tissue, the otolithic membrane connects to the tops of the hair bundles and provides mechanical coupling between them. This inter-cell coupling is sufficiently strong to suppress the innate motility of individual bundles, poising the full coupled system in the quiescent regime. Only after digestion and careful removal of this membrane have spontaneous hair bundle oscillations been observed. In this study, we explore the possibility that MD results from degraded or missing otolithic membrane tissue in the utricle or saccule. Specifically, we hypothesize that locally degraded coupling between hair cells can lead to the sudden onset of spontaneous oscillations in a subset of bundles, leading to a spurious signal that may cause VDAs and the vestibular dysfunction seen in patients with MD.

We use a previously developed numerical model to describe the nonlinear dynamics of an array of active hair cells. We introduce elastic coupling between nearest and next-nearest neighbors on the grid, representing the mechanical connection between cells imposed by the otolithic membrane tissue. We introduce heterogeneity in the selection of the model parameters to produce spatially random dispersion in the characteristic frequencies of the hair cells, approximating that of the saccule. This frequency dispersion suppresses the autonomous motion of the hair bundles, resulting in a quiescent system. We then reduce the coupling strength of a selected region of the membrane to model the effects of tissue damage. We explore the dynamics for several levels of tissue damage and find that large, abrupt spikes in hair bundle position emerge, with larger and more frequent spikes occurring for increasing levels of tissue damage. These spikes in bundle position correspond to large spikes in the opening probability of the transduction channels, which would hence elicit significant neural activity. We therefore propose that aspects of the vestibular disfunction attributed to Meniere's Disease arise from degradation in the otolithic membrane tissue, which reduces the coupling strength imposed by the membrane, resulting in large, abrupt spikes in the positions of the hair bundles.

**METHODS:**

To explore the impact of Meniere's disease in a numerical simulation, we model the sacculus as a system of coupled nonlinear oscillators. The mechanical motility of individual hair bundles is described using previously established biophysical models, which have been shown to reproduce a broad set of experimental results [3,12]. We model the coupled system as a 15 by 15 grid of hair cells, with each cell connected to its nearest and next-nearest neighbors (Fig. 1). To simulate localized damage to the otolithic membrane, we impose varying levels of reduction in the coupling coefficient, $K_d$, within the damaged region (shown in red in Figure 1). Outside of the damaged region, the coupling constants, K, are held constant to represent healthy otolithic membrane tissue.

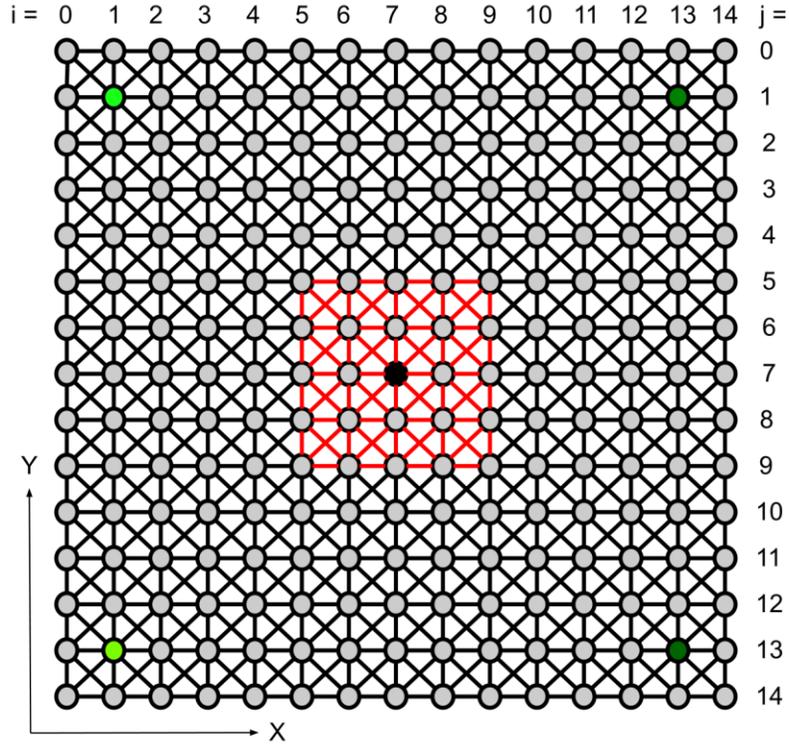

*Figure 1: Grid of 15 x 15 coupled hair bundles, each represented by a circle. The black lines represent springs with coupling constant, K, and red lines represent springs with coupling constant, $K_d$, which are varied to simulate damage to the membrane. Green circles represent the bundles used to compare the dynamics of bundles under normal coupling conditions to those with a damaged membrane. The black circle represents the bundle in the damaged region that will be examined.*

The equations of motion for the bundles of each individual hair cell are as follows:

$$\frac{\lambda}{a_0[i,j]} \frac{dX}{dt} = -K_{gs}(X - X_a - DP_o) - K_{sp}X + F_K + \eta$$

$$\frac{\lambda_a}{a_0[i,j]} \frac{dX_a}{dt} = K_{gs}(X - X_a - DP_o) - \gamma N_a f[i,j] p(C) + K_{es}(X_a + X_{es}) + \eta_a$$

$$\frac{\tau}{a_0[i,j]} \frac{dC}{dt} = C_0 - C + C_m P_o + \delta c.$$

X represents the position of the hair bundle, $X_a$ represents the position of the myosin motor, and C represents the $Ca^{2+}$ concentration at the myosin motor binding site. The hair bundles are restricted to one dimensional motion along the x-axis, as defined in Figure 1. Each hair bundle's transduction channel is viewed as having two states, open or closed, with open probability,

$$P_o = \frac{1}{1 + Ae^{-(X-X_a)/\delta}},$$

where $A = \exp([\Delta G + (K_{gs}D^2)/(2N)]/(K_bT))$,

and $\delta = NK_bT/(K_{gs}D)$.

The force generated by a single myosin motor is given by $f[i,j] = f_{max}[i,j] / (N_a P_o)$, where each element of $f_{max}[i,j]$ is randomly sampled from a uniform distribution ranging from 87 pN to 352 pN. The unitless feedback parameter, S, and the maximal force $f_{max}$ define the dynamic state of the hair bundle, poising it in the oscillatory, bistable, or quiescent regime, or near a bifurcation between the different regions. Prior studies have established a full state diagram, using this model, and defined the boundaries delineating these regions. Two specific points were identified within the diagram: the point at which a bundle would exhibit the greatest sensitivity, as well as a point that best describes the experimentally determined state; both points reside within the oscillatory region of the state diagram [12].

We modify the model of Nadrowski et al to include an additional term in the equation of motion for the myosin motors. The term was introduced in prior literature to describe the experimental observation that myosin mediated adaptation is not complete [8, 20]. This effect is represented by an additional elastic spring, which opposes adaptation. $K_{es}$ represents the coefficient of this extension spring, and $X_{es}$ describes the position of the spring relative to the myosin motors [16].

To mimic the natural variation of hair bundle states in a sacculus, we vary the parameters in the numerical model of each oscillator, thus leading to a dispersion in their innate frequencies (Fig. 2). Further, the selection of parameters leads to a variation in the dynamic state exhibited by each bundle. As an approximation of this dispersion, we assume that they are all poised in the oscillatory state, and sample the position of each bundle uniformly along the line segment described as follows. The probability that myosin motors are bound to actin is given by $p(C) = p_0 + p_1[i,j]C$. $p_0$ is fixed to 0.2 and each $p_1$ value is sampled in such a manner that the dimensionless feedback parameter S is linearly related to $f_{max}$. We have obtained similar results when poising all of the oscillators in regions of the oscillatory region closer to the Hopf bifurcation.

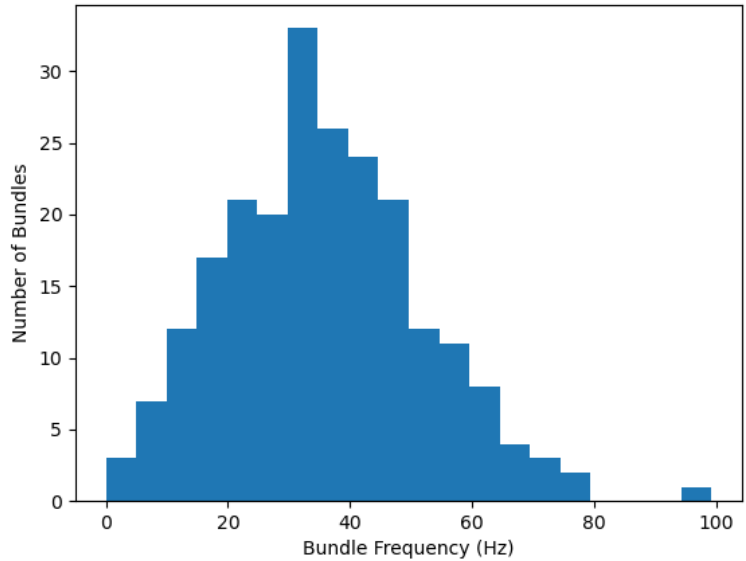

*Figure 2: Frequency distribution of 15 x 15 uncoupled hair bundles. The $a_0$, S, and $f_{max}$ values were randomly selected for each hair bundle.*

Prior experimental studies have shown that hair cells of an amphibian sacculus show a random distribution of innate oscillation frequencies. We assume a comparable distribution in our model and assign each hair cell a characteristic $a_0$, randomly selected from a normal distribution of mean 1.5 and standard deviation 0.25. These $a_0$ values ensure that the frequency distribution of uncoupled hair bundles in our model matches the measured frequency distribution of spontaneous oscillations in the bullfrog sacculus [14]. Each hair bundle thus has its own individual values of $a_0$, S, and $f_{max}$ that remain fixed for the duration of each trial.

Each equation is subject to white Gaussian noise, with zero mean and correlation functions as follows:

$$<\eta(t)\eta(t')> = 2K_b T \lambda \delta(t-t')$$

$$<\eta_a(t)\eta_a(t')> = 2K_b T_a \lambda_a \delta(t-t')$$

$$<\delta c(t)\delta c(t')> = 2C_M^2 N^{-1} P_o(1-P_o)\tau_C \delta(t-t'),$$

where the angle brackets denote the time average and $\delta$ represents the Dirac delta function. The force exerted on each bundle due to coupling to nearest and next-nearest neighbors is given by

$$F_k(K_d) = K_v \left(1 - \frac{L_0}{\sqrt{(\Delta X + kd)^2 + (ld)^2}}\right)(\Delta X + kd).$$

where $\Delta X$ is the difference in the positions of the two bundles connected by the coupling spring, and $L_0 = \sqrt{(k^2 + l^2)d^2}$. d is the distance between nearest neighbors, while k and l are integers that represent the relative position between bundles on the grid. k is the row of the first bundle minus the row of the second, and l is the column of the first bundle minus the column of the second. This notation accounts for the fact that nearest neighbors are closer together than next-nearest neighbors.

$K_v = K$ in healthy tissue and $K_v = K_d$ in damaged tissue, as shown in figure 1. We vary $K_d$ in order to explore the dynamics of the coupled system for different degrees of damage to the otolithic membrane tissue. The coupled differential equations were solved numerically using a fourth order Runge-Kutta method with time steps of 20 μs. We note that the original model contains offsets in X, $X_a$, and C; as these offsets reflect an arbitrary choice of a zero point and do not affect the bundle dynamics, we have elected to retain them in our adaptation of this model [3]. The values of all constants can be found in Table 1.

| Constant | Value | Description |
|---|---|---|
| $\lambda$ | $2.8 * 10^{-6}$ Ns/m | *Friction coefficient of hair bundle* |
| $\lambda_a$ | $10 * 10^{-6}$ Ns/m | *Friction coefficient of adaptation motors* |
| $K_{gs}$ | $750 * 10^{-6}$ N/m | *Combined gating spring stiffness* |
| $K_{sp}$ | $600 * 10^{-6}$ N/m | *Combined stiffness of stereociliary pivots and load* |
| $K_{es}$ | $140 * 10^{-6}$ N/m | *Stiffness of extent spring* |
| $x_{es}$ | $20 * 10^{-9}$ m | *Resting deflection of extension spring* |
| dgs | $8.7 * 10^{-9}$ m | *Gating-spring elongation on channel opening* |
| $\gamma$ | 0.14 | *Geometrical gain of stereociliary shear motion* |
| $\tau$ | $0.1 * 10^{-3}$ s | *Time constant of calcium feedback* |
| $\tau_c$ | $1 * 10^{-3}$ s | *Dwell time of transduction channels* |
| $C_0$ | 0 M | *Intracellular Ca2+ concentration with closed channels* |
| N | 50 | *Number of stereocillia* |
| $N_a$ | 3000 | *Number of motors in the hair bundle* |
| $K_b$ | $1.38 * 10^{-23}$ J/K | *Boltzmann constant* |
| T | 300 K | *Temperature* |
| $\Delta G$ | $4.14 * 10^{-20}$ J | *Gibbs free energy* |
| $T_a$ | $1.5 * T$ | *Effective temperature for noise strength of motor position* |
| D | dgs/gamma | *Displacement due to gating spring* |
| $C_{max}$ | $250 * 10^{-3}$ M | *Maximum calcium concentration at motor* |
| d | $50 * 10^{-6}$ m | *Distance between hair bundles* |
| K | .0014 N/m | *Combined stiffness of the membrane, hair bundles, and filaments* |

*Table 1: Parameters of the numerical model.*

**RESULTS:**

When autonomous oscillators with significant frequency dispersion are strongly coupled, the oscillations may become suppressed through a phenomenon known as amplitude death. It has previously been proposed that this mechanism is responsible for quenching the spontaneous hair-bundle oscillations in the sacculus, an organ which possesses both strong coupling and significant frequency dispersion [6]. Experimental studies have further shown that the presence of a healthy otolithic membrane suppresses innate oscillations exhibited by uncoupled hair bundles. Our model exhibits amplitude death when $K_v = K_d = K$ for all oscillators, describing a properly functioning sacculus. However, as the coupling coefficient decreases, the suppression weakens, and bundles begin to exhibit spontaneous activity with amplitudes significantly greater than those induced by the thermal noise alone (Fig. 3). These spikes in amplitude increase in both magnitude and frequency with reduction of coupling, which models increasing levels of damage to the tissue.

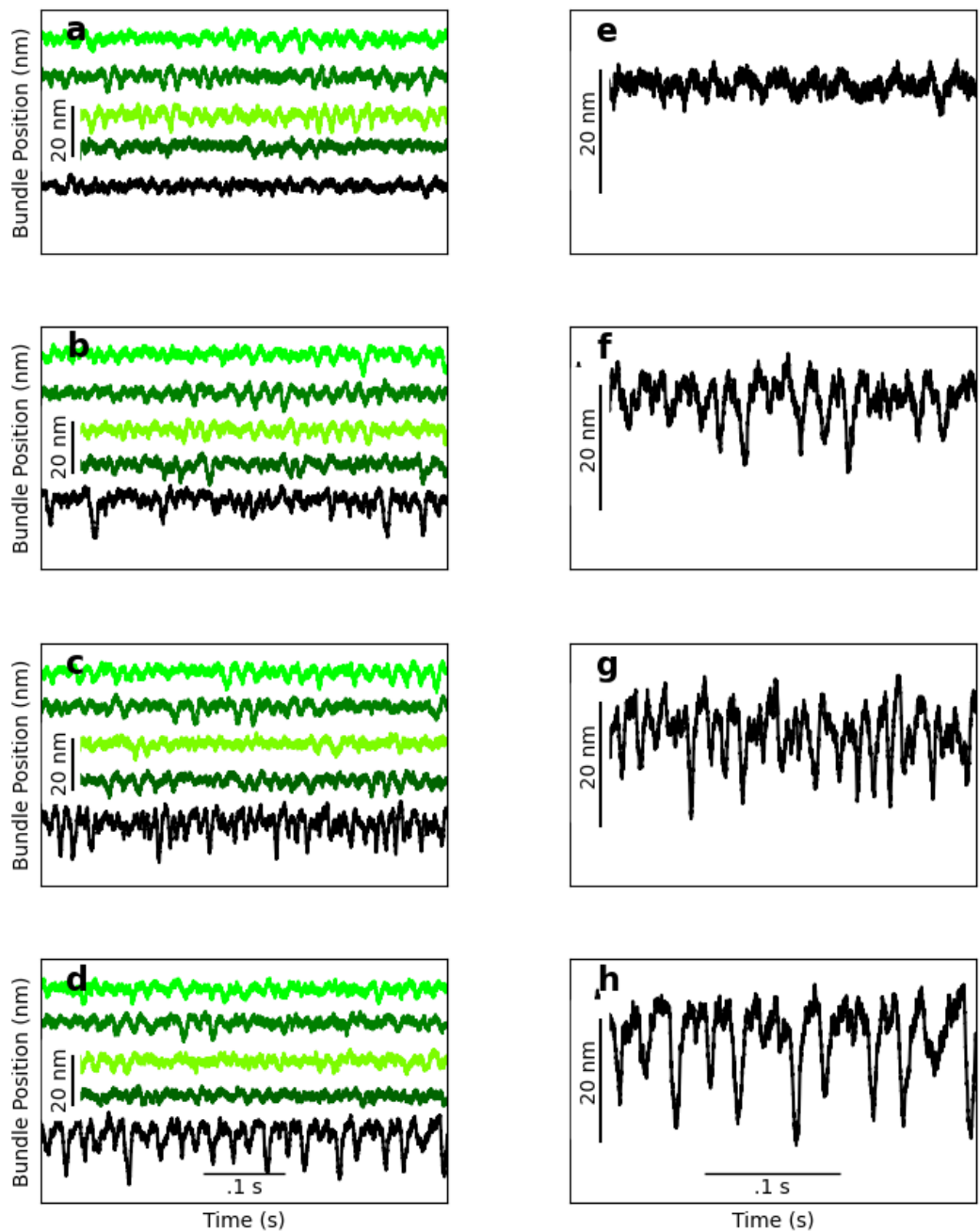

*Figure 3: (a-d) Traces of bundle position with $K_d$ = K, K/5, K/10, and K/15, respectively. The colors of each trace represent the colored bundles in figure 1. The black bundle resides in damaged tissue while the rest reside in healthy tissue. Arbitrary offsets have been added to the traces for clarity. (e-h) Zoomed in plots of the black traces in a-d.*

We present these statistics in histograms, which display long tails in the negative direction, becoming more skewed with increasing level of damage (Fig. 4). These drastic spikes in bundle positions from equilibrium have been shown to translate into a change in the opening probability of the transduction channels, which in turn leads to depolarization of the hair cell soma that triggers neural activity [7, 13, 19]. These results suggest that lowered coupling strength due to otolithic membrane damage, degradation, or absence may lead to the vestibular dysfunction described by Meniere's Disease.

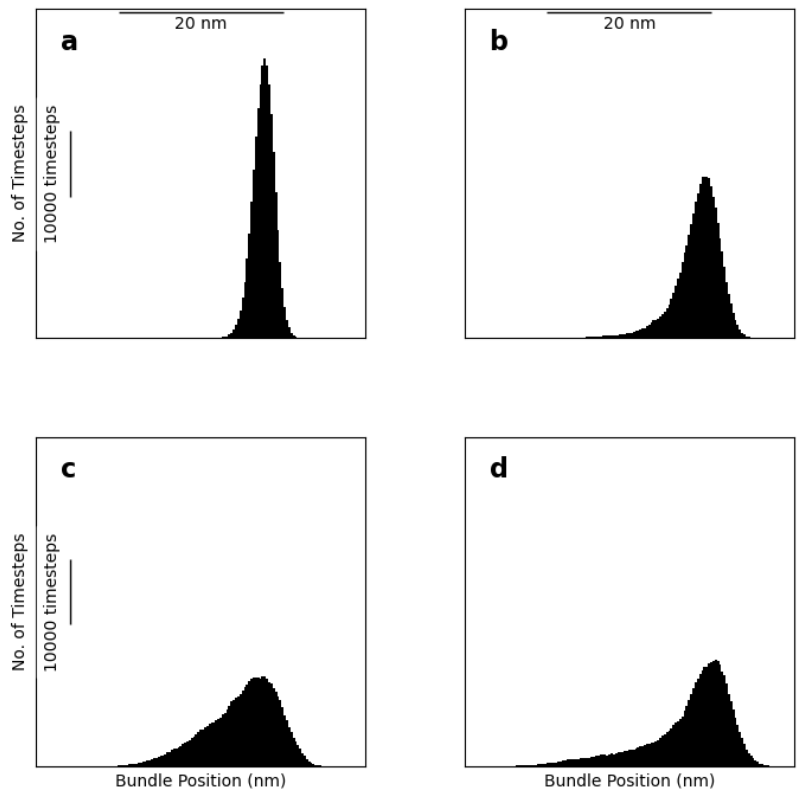

*Figure 4: (a-d) Histograms of position of a bundle under damaged conditions for $K_d = K$, $K/5$, $K/10$, and $K/15$, respectively. The traces used to generate these histograms are plotted in black in Fig. 3. Each histogram was produced using a trace from a simulation of 500,000 timesteps representing 10 seconds.*

## DISCUSSION:

Our numerical simulations show that as the coupling strength imposed by the otolithic membrane decreases, hair bundles exhibit abrupt spikes and even intermittent oscillations in their position. This spontaneous activity translates to large fluctuations in the open probability of the transduction channels. Prior work has shown that channel opening leads to depolarization of the hair cell soma and could result in spurious signal detection by the afferent neurons. These results agree with the surgical observation that vestibular drop attacks are correlated with significant damage and degradation of the otolithic membrane in patients with Meniere's Disease. Additionally, our findings explain why the absence of a functioning gene for otogelin, an inner-ear structural protein that attaches the otolithic membrane to the underlying tissue, can lead to MD [15, 18]. When functional otogelin proteins are not available, the otolithic membrane is not securely attached to the neuroepithelia, and in some cases completely detaches. Hence, the extant literature indicates that reduction or elimination of innate coupling between hair cells underlies a number of symptoms associated with the MD.

It has previously been proposed that the vestibular dysfunction associated with Meniere's Disease may result from the altered fluid dynamics imposed by endolymphatic hydrops [5, 17]. While both effects may be present in patients suffering from MD, in this study, we focus on a theoretical description of how these symptoms can arise purely from the degradation of the otolithic membrane tissue without the need for forces caused by endolymphatic hydrops. We use a simple numerical model that has previously been developed to describe the active, nonlinear dynamics of hair cells. Although this model was originally proposed to describe the spontaneous oscillations of saccular hair-cell bundles in the bullfrog, the finding that reduced coupling can lead to spontaneous spikes in hair bundle position is very general. Any inter-connected active system in which coupling between the nonlinear elements leads to quiescence can exhibit sudden motion when localized damage removes the quenching of active motility. We hence show a possible mechanism by which degradation of coupling can lead to spurious vestibular signals.

Future work entails using animal models to test for this mechanism of spurious detection in groups of active hair cells. The hair-cell bundles can be coupled with artificial membranes of different size, thickness, and properties of the material, to explore the dynamics of various coupled active systems. Alternatively, semi-intact preparations can be used, in which the otolithic membrane is left attached to the hair bundle tips, and one can impose various degrees of damage to the tissue, thus mimicking the MD symptoms *in vitro*. Future computational work entails exploring the effects of various shapes, sizes, and levels of heterogeneity of the damaged region to determine if this model can encompass the full range of observed phenomena associated the MD.

**CONCLUSIONS:**

The results of our numerical simulations suggest that damage to the otolithic membrane of the utricle and/or saccule may be sufficient to induce vestibular drop attacks and the vestibular dysfunction seen in patients with Meniere's Disease. These observations agree with prior work showing that patients with damaged/dysfunctional otolithic membranes in the utricle and/or saccule often develop MD, and some experience VDAs. Future experimental work is needed to confirm these results *in vitro*, and future computational work is needed to determine if the effects of reduced coupling can explain other aspects of Meniere's Disease.

**REFERENCES:**

[1] Benser, M. E., Marquis, R. E. & Hudspeth, A. J. Rapid, active hair bundle movements in hair cells from the bullfrog's sacculus. *Journal of Neuroscience* 16, 5629–5643 (1996).

[2] Calzada, A. P., Lopez, I. A., Ishiyama, G. & Ishiyama, A. Otolithic membrane damage in patients with endolymphatic hydrops and drop attacks. *Otology and Neurotology* 33, 1593–1598 (2012).

[3] Dierkes, K., Jülicher, F. & Lindner, B. A mean-field approach to elastically coupled hair bundles. *European Physical Journal E* 35, (2012).

[4] Fredrickson-Hemsing, L., Strimbu, C. E., Roongthumskul, Y. & Bozovic, D. Dynamics of freely oscillating and coupled hair cell bundles under mechanical deflection. *Biophysical Journal* 102, 1785–1792 (2012).

[5] Ishiyama, G., Lopez, I. A., Sepahdari, A. R. & Ishiyama, A. Meniere's disease: Histopathology, cytochemistry, and imaging. *Annals of the New York Academy of Sciences* 1343, 49–57 (2015).

[6] Kim, K. J. & Ahn, K. H. Amplitude death of coupled hair bundles with stochastic channel noise. *Physical Review E - Statistical, Nonlinear, and Soft Matter Physics* 89, (2014).

[7] LeMasurier, M. & Gillespie, P. G. Hair-cell mechanotransduction and cochlear amplification. *Neuron* vol. 48 403–415 (2005).

[8] Martin, P., Bozovic, D., Choe, Y. & Hudspeth, A. J. Spontaneous oscillation by hair bundles of the bullfrog's sacculus. *Journal of Neuroscience* 23, 4533–4548 (2003).

[9] Martin, P. & Hudspeth, A. J. Active hair-bundle movements can amplify a hair cell's response to oscillatory mechanical stimuli. *Proceedings of the National Academy of Sciences of the United States of America* 96, 14306–14311 (1999).

[10] Martin, P., Hudspeth, A. J. & Jülicher, F. Comparison of a hair bundle's spontaneous oscillations with its response to mechanical stimulation reveals the underlying active process. *Proceedings of the National Academy of Sciences of the United States of America* 98, 14380–14385 (2001).

[11] Martin, P. & Hudspeth, A. J. Compressive nonlinearity in the hair bundle's active response to mechanical stimulation. *Proceedings of the National Academy of Sciences of the United States of America* 98, 14386–14391 (2001).

[12] Nadrowski, B., Martin, P. & Jülicher, F. Active hair-bundle motility harnesses noise to operate near an optimum of mechanosensitivity. *Proceedings of the National Academy of Sciences of the United States of America* 101, 12195–12200 (2004).

[13] Ó Maoiléidigh, D. & Ricci, A. J. A Bundle of Mechanisms: Inner-Ear Hair-Cell Mechanotransduction. *Trends in Neurosciences* vol. 42 221–236 (2019).

[14] Ramunno-Johnson, D., Strimbu, C. E., Fredrickson, L., Arisaka, K. & Bozovic, D. Distribution of frequencies of spontaneous oscillations in hair cells of the bullfrog sacculus. *Biophysical Journal* 96, 1159–1168 (2009).

[15] Roman-Naranjo, P. *et al.* Burden of rare variants in the OTOG Gene in familial meniere's disease. *Ear and Hearing* 1598–1605 (2020).

[16] Roongthumskul, Y., Fredrickson-Hemsing, L., Kao, A. & Bozovic, D. Multiple-timescale dynamics underlying spontaneous oscillations of saccular hair bundles. *Biophysical Journal* 101, 603–610 (2011).

[17] Sajjadi, H. & Paparella, M. M. Meniere's disease. *The Lancet* vol. 372 406–414 (2008).

[18] Simmler, M. C. *et al.* Targeted disruption of Otog results in deafness and severe imbalance. *Nature Genetics* 24, 139–143 (2000).

[19] Vollrath, M. A., Kwan, K. Y. & Corey, D. P. The micromachinery of mechanotransduction in hair cells. *Annual Review of Neuroscience* vol. 30 339–365 (2007).

[20] Yamoah, E. N. & Gillespie, P. G. Phosphate analogs block adaptation in hair cells by inhibiting adaptation-motor force production. *Neuron* 17, 523–533 (1996).